\documentclass[screen, manuscript, natbib=true, anonymous=false, sigconf]{acmart}
\AtBeginDocument{%
  }
    \usepackage{tikz}
\usetikzlibrary{positioning, arrows.meta}
\usepackage{tikz}
\usetikzlibrary{shapes.geometric, arrows.meta, positioning, fit, calc, backgrounds, shadows, patterns}

\definecolor{corpblue}{RGB}{235, 245, 255}    
\definecolor{darkblue}{RGB}{45, 85, 125}      
\definecolor{corpred}{RGB}{255, 240, 240}     
\definecolor{darkred}{RGB}{145, 45, 45}       
\definecolor{slate}{RGB}{240, 242, 245}       
\definecolor{charcoal}{RGB}{70, 75, 80}       
\usepackage{graphicx}  
\usepackage[table]{xcolor} 
\usepackage{booktabs}  
\usepackage[most]{tcolorbox}
\usepackage{fontawesome5} 
\begin{document}

\title{Mitigating Preference Leakage via Strict Estimator Separation for Normative Generative Ranking}

\author{Dalia Nahhas}
\affiliation{
  \institution{University of Southampton}
  \country{United Kingdom}
}
\affiliation{
  \institution{Umm Al-Qura University}
  \country{Saudi Arabia}
}
\email{D.G.Nahhas@soton.ac.uk}

\author{Xiaohao Cai}
\affiliation{
  \institution{University of Southampton}
  \country{United Kingdom}
}
\email{X.Cai@soton.ac.uk}

\author{Imran Razzak}
\affiliation{
  \institution{Mohamed bin Zayed University of Artificial Intelligence}
  \country{United Arab Emirates}
}
\email{imran.razzak@mbzuai.ac.ae}

\author{Shoaib Jameel}
\affiliation{
  \institution{University of Southampton}
  \country{United Kingdom}
}
\email{M.S.Jameel@southampton.ac.uk}

\begin{abstract}
In Generative Information Retrieval (GenIR), the bottleneck has shifted from generation to the selection of candidates, particularly for normative criteria such as cultural relevance. Current LLM-as-a-Judge evaluations often suffer from circularity and preference leakage, where overlapping supervision and evaluation models inflate performance. We address this by formalising cultural relevance as a within-query ranking task and introducing a leakage-free two-judge framework that strictly separates supervision (Judge B) from evaluation (Judge A). On a new benchmark of 33,052 (NGR-33k) culturally grounded stories, we find that while classical baselines yield only modest gains, a dense bi-encoder distilled from a Judge-B-supervised Cross-Encoder is highly effective. Although the Cross-Encoder provides a strong supervision signal for distillation, the distilled BGE-M3 model substantially outperforms it under leakage-free Judge~A evaluation. We validate our framework on the human-curated Moral Stories dataset, showing strong alignment with human norms. Our results demonstrate that rigorous evaluator separation is a prerequisite for credible GenIR evaluation, proving that subtle cultural preferences can be distilled into efficient rankers without leakage. \textit{[Code and experimental settings: Anonymous] [Dataset: Anonymous]}.
\end{abstract}



\maketitle

\section{Introduction}
\label{sec:introduction}

Generative Information Retrieval (GenIR) increasingly supports systems that combine retrieval, generation, and downstream decision making, including retrieval-augmented generation and creative text applications~\cite{lewis2020retrievalaugmented,Alaofi2024generativeinformation,bonifacio2022inpars}. A common operating mode is \emph{generate--then--select}: given the same constraints, a model produces multiple fluent candidates, and system quality depends primarily on \emph{selection}---choosing the best candidate among many possible alternatives. Throughout this work, ``within-query'' refers to ranking candidates generated under identical constraint tuples; candidates from different queries are never compared.

Selection becomes especially difficult when the target criterion is \emph{normative} rather than purely topical. In a culturally grounded generation, candidates may be equally fluent and on-topic while differing in whether they express culturally specific settings, practices, values, or communication style. Standard lexical signals and generic dense similarity are not designed to capture such \emph{non-topical} relevance dimensions, making within-query ordering unreliable under these criteria~\cite{manning2008ir,jarvelin2002cumulated}. This setting yields candidate pools that are tightly matched in topical relevance and coherence, while differing primarily along \emph{normative cultural grounding}, a latent dimension beyond surface lexical overlap. As illustrated in Figure~\ref{fig:motivating_example}, standard lexical rankers often struggle to reliably distinguish between surface-level stereotypes (e.g., hummus, falafel) and deep normative grounding (e.g., the specific etiquette of refusal and insisting), a gap our framework aims to bridge.

\begin{figure}[t] 
\begin{tcolorbox}[
    enhanced, colback=white, colframe=charcoal, boxrule=0.5pt, arc=2mm, drop shadow,
    left=3pt, right=3pt, top=3pt, bottom=3pt, 
    title={\small \textbf{\textsc{Example:} Surface vs. Normative Relevance} \\ \quad \footnotesize $\langle$ Age: 10; Moral: Generosity; Culture: Arab $\rangle$},
    fonttitle=\sffamily\bfseries, coltitle=white, boxed title style={colback=charcoal}
]
    \footnotesize 
    \textbf{\color{darkred}\faTimesCircle\ Candidate A (Surface):} 
    \itshape ``Ahmed sat in the \textbf{desert} eating \textbf{hummus}... gave him a \textbf{high-five}.'' 
    \upshape \hfill \textit{$\rightarrow$ Stereotypical keywords, western norms.}
    
    \vspace{0.3em}
    \textbf{\color{darkblue}\faCheckCircle\ Candidate B (Normative):} 
    \itshape ``Ahmed \textbf{insisted}... guest \textbf{refused three times}... \textbf{piled more lamb}...'' 
    \upshape \hfill \textit{$\rightarrow$ Captures the ``ritual of refusal''.}
    
    \vspace{0.3em}
    \textbf{Gap:} Lexical rankers prefer A (keywords); Normative rankers must identify B (behavior).
\end{tcolorbox}
\vspace{-0.08in}
\caption{An example of a query where dense bi-encoders and lexical models can struggle. They rank the stereotypical Candidate A higher due to token overlap, whereas the true cultural ground truth (Candidate B) requires reasoning about the enactment of the moral value.}
\label{fig:motivating_example}
\vspace{-1.3em} 
\end{figure}

To scale evaluation of such criteria, recent work increasingly uses large language models (LLMs) as rubric-guided judges~\cite{zheng2023mtbenchjudge,Hashemi2024llmrubric,li2024llmsasjudgescomprehensive}. However, judge-based pipelines are vulnerable to \emph{circular evaluation} \cite{dietz2025principles}: if the same judge (or model family) provides both supervision and evaluation, reported gains can reflect evaluator preference rather than real improvements. This risk is amplified by \emph{self-preference bias} \cite{chen2025beyond}, where judges systematically favour outputs from their own model family, leading to \emph{preference leakage} \cite{yang2025code} and inflated ranking estimates under insufficient separation~\cite{koo2024benchmarkingcognitive,chen2024humansllmsjudgestudy,li2025preferenceleakage}.

We address these challenges by casting culturally grounded story selection as a \emph{within-query graded ranking} task and introducing a \emph{two-judge leakage-free framework}. Judge~B provides rubric scores used \emph{only} for ranking signals and training, while an independent Judge~A supplies labels used \emph{only} for final evaluation. To ensure fair comparison across methods, we evaluate on identical within-query candidate pools defined by the intersection of valid annotations from both judges. This design avoids missing-label confounds and ensures that all methods are evaluated on exactly the same candidate set per query. Our results show that rubric-based cultural judgments remain informative under an independent evaluator, and that this signal can be distilled \cite{song2025self} into neural ranking models without inference-time judging. We validate the judges' alignment with human norms using the Moral Stories \cite{emelin2021moral} benchmark. Unlike prior LLM-as-a-judge \cite{gu2024survey} setups that conflate supervision and evaluation, our framework enforces \emph{role separation} and evaluates all methods under a fixed intersection pool, enabling leakage-free, within-query ranking comparisons.

\emph{\textbf{Contributions:}} We formalise culturally grounded selection as a conditional ranking optimisation problem, where the objective is to learn a scoring function $f_\theta(s\text{: story},q \text{: query})$ that approximates a latent normative prior $R^*$ within dense, constraint-matched candidate pools. We introduce a ``strictly orthogonal two-judge'' framework ($J_{\rm supervision} \cap J_{\rm evaluation} = \emptyset$) that eliminates circular preference leakage. By enforcing evaluation exclusively on the unified intersection pool $\mathcal{S}_q^{\cap}$, we mathematically control for estimator availability bias and ensure valid within-query variance. We empirically identify a Normative Feature Suppression bottleneck in direct bi-encoder training, demonstrating that normative relevance features are latent and overshadowed by the dominant semantic components of pre-trained embeddings. We find that directly fine-tuning a bi-encoder under Judge-B normative supervision can be unstable, whereas interaction-based cross-encoders provide a strong, high-capacity teacher reference. We further show that a teacher–student distilled bi-encoder transfers much of this signal and achieves strong performance while retaining bi-encoder effectiveness. We release NGR-33k (Normative Generative Ranking dataset), a benchmark for complete reproducibility artefacts. We validate our framework's external validity on human-curated data (Moral Stories), achieving strong inter-judge alignment.

\vspace{-3mm}

\section{Related Work}
\label{sec:related_work}

\emph{\textbf{Retrieval vs.\ Selection}}: GenIR integrates generation and retrieval and has become central to modern retrieval-augmented pipelines~\cite{lewis2020retrievalaugmented,Alaofi2024generativeinformation}. Within this space, prior work can be broadly divided into: i) \emph{model-based retrieval} that replaces explicit indexing with parametric access, and ii) \emph{generate-then-rank} pipelines that explicitly construct candidate sets and then select among them. 

Model-based retrieval approaches, such as Differentiable Search Indices (DSI), aim to internalise the corpus directly into model parameters, enabling retrieval via generation rather than explicit indexing~\cite{tay2022transformermemory,bevilacqua2022autoretrieval}. While these approaches emphasise memorisation and fully differentiable retrieval, they are not designed for the \emph{within-query} selection problem central to our work: ordering several candidates generated under the same query constraints. They also do not engage with leakage issues that arise when rubric-based supervision and evaluation are both provided by LLM judges.

In contrast, generate-then-rank pipelines~\cite{bonifacio2022inpars,lewis2020retrievalaugmented} shift the bottleneck to \emph{post-generation selection}, where candidates produced via stochastic decoding~\cite{holtzman2020curiouscaseneuraltext} require rigorous sorting frameworks~\cite{Alaofi2024generativeinformation}. We explicitly target this setting by formulating culturally grounded storytelling as a \emph{within-query graded ranking} task evaluated with standard IR measures~\cite{manning2008ir,jarvelin2002cumulated,chapelle2009expected,mitra2018introduction}, distinguishing normative grounding from simple fluency.

\emph{\textbf{LLMs as Judges and Evaluation Biases}}: LLMs have recently been adopted as automatic evaluators for open-ended generation tasks \cite{arabzadeh2025human}, often showing strong correlation with human judgments under carefully designed rubrics~\cite{zheng2023mtbenchjudge,Hashemi2024llmrubric,li2024llmsasjudgescomprehensive}. However, judge-based evaluation is sensitive to prompting, decoding, and judge choice, and can exhibit systematic biases; this motivates uncertainty-aware reporting and framework constraints~\cite{chang2024surveyllmevaluation,yamauchi2025empiricalstudyllm,leaderboard2024judge,guo2023evaluatinglargelanguagemodels,zhao2025surveylargelanguagemodels}.

A particularly concerning phenomenon is \emph{self-preference bias}, where LLM judges unreasonably favour outputs generated by models from their own family or with similar training distributions~\cite{koo2024benchmarkingcognitive,chen2024humansllmsjudgestudy}. This bias is especially acute in within-query ranking scenarios, where candidates are highly similar and small scoring deviations can substantially alter rankings. Importantly, self-/family preference is a \emph{bias} of the evaluator, while leakage is a \emph{framework confound}: using overlapping signals for training and evaluation can make improvements appear even when they do not transfer to an independent evaluator. \citet{li2025preferenceleakage} formalise this issue as \emph{preference leakage}, showing that overlapping supervision and evaluation signals can lead to inflated and misleading performance estimates.

Existing mitigation strategies include cross-model judging, ensemble evaluation, and prompt randomization~\cite{chen2024humansllmsjudgestudy}. Our work complements these approaches by enforcing strict role separation between supervision and evaluation through a two-judge framework with blocked information flow, and by enforcing identical candidate pools via intersection filtering to remove \emph{control} missing-label confounds in comparisons~\cite{voorhees2000trec,smucker2007comparison,carterette2012multipletesting,demsar2006statistical}.

\emph{\textbf{Cultural Relevance (Non-Topical Relevance)}}: Cultural aspects of language have been extensively studied in the context of bias detection, fairness, and harm mitigation, where culture is often treated as an attribute to control, measure, or remove~\cite{blodgett2020language,huang2023culturalbias,tao2024culturalbiasalignment,alkhamissi2024investigatingcultural,naous2024havingbeerprayermeasuring,sukiennik2025evaluationcultural}. Related benchmark efforts likewise implement social structure attributes primarily as risks to surface or mitigate~\cite{parrish2022bbq,Dhamala2021BOLD,smith2022imsorryhearthat,bender2021dangers}. Policy and governance perspectives further warn of a \emph{cultural compatibility gap} and \emph{algorithmic monoculture}, where models default toward dominant norms absent neutralising constraints~\cite{adalovelace2025tokenisingculture}.

While benchmarks such as CulturalBench~\cite{chiu2025culturalbench} and NormAd~\cite{rao2025normad} probe factual knowledge or social acceptability, they do not address the generative selection of fully formed narratives. Our NGR-33k fills this gap by treating culture not as a bias axis, but as a positive, graded relevance signal~\cite{manning2008ir,jarvelin2002cumulated} for ranking ``thick'' normative enactment, measurable via standard IR metrics~\cite{chapelle2009expected}.

\emph{Culturally grounded children's story generation:} Research on culturally grounded narratives indicates that implicit cues require treating culture as an explicit ranking objective~\cite{Zhang2022storybuddy,Ye2025colin,rezapouretal2025tales,rooein2025biasedtalesculturaltopic}. While normative resources support supervision beyond surface similarity~\cite{forbes2020socialchemistry,emelin2021moral}, rubric-based judging offers a scalable proxy for ``thick evaluation''~\cite{qadri2025casethickevaluationscultural}. Unlike work framing culture as alignment diagnosis~\cite{masoud2024culturalalignment,li2024culturellm,choenni2025selfalignment}, we focus on explicit within-query selection.

\emph{Benchmark gap for culturally grounded story selection:} To our knowledge, no benchmark jointly supports age, moral, and culturally grounded \emph{within-query ranking} under strict evaluator separation. While narrative datasets focus on coherence~\cite{mostafazadeh2016storycloze,hill2016cbt} and cultural corpora offer broad coverage~\cite{li2024ssgen,hagedorn2022folktales}, none enforce this constraint triad within a fixed candidate-pool framework, motivating our controlled generate-then-rank design.

\emph{\textbf{Learning to Rank and Ranking Distillation}}: Learning-to-rank (LTR)~\cite{liu2009learning} optimizes ordering using graded signals~\cite{reddy2024first,burges2005ranknet,dehghani2017neuralranking,burges2010ranknetlambdarank}. We employ ranking distillation \cite{tang2018ranking} to transfer expensive normative supervision~\cite{Hashemi2024llmrubric,li2024llmsasjudgescomprehensive} into lightweight models~\cite{zhou2018weakly}, recovering the signal without inference-time costs. However, this is only valid if improvements persist under independent evaluation, mitigating evaluator-family preference~\cite{koo2024benchmarkingcognitive,chen2024humansllmsjudgestudy,li2025preferenceleakage}.

\emph{\textbf{Data and Query Construction in GenIR}}: Several recent works construct structured or synthetic queries to study generative retrieval and ranking \cite{wu2024generative} under controlled conditions, enabling reproducible evaluation and clearer attribution of ranking gains~\cite{bonifacio2022inpars,Alaofi2024generativeinformation}. Such designs reduce topical distraction and enable analysis of \emph{within-query} ranking behaviour, where candidates share constraints and are intentionally similar.

Our query construction follows this line of work by defining structured constraint triples (age, moral, culture) that induce candidate pools with shared topical content. This setup isolates the selection problem by ensuring that observed differences among candidates primarily reflect normative and stylistic factors rather than topical mismatch, and supports query-level evaluation using graded relevance metrics~\cite{manning2008ir,jarvelin2002cumulated,chapelle2009expected}.

\emph{\textbf{POSITIONING.}} While GenIR identifies the post-generation selection bottleneck~\cite{Alaofi2024generativeinformation,lewis2020retrievalaugmented,bonifacio2022inpars} and LTR provides optimization mechanisms~\cite{liu2009learning,burges2005ranknet,dehghani2017neuralranking,zhou2018weakly}, current approaches overlook a critical validity crisis: judges are plagued by instability and preference leakage~\cite{koo2024benchmarkingcognitive,chen2024humansllmsjudgestudy,li2025preferenceleakage,blackwell2025reproduciblellm,liang2023helm,huang2024trustllm}, and culture is reduced to a bias-mitigation task~\cite{blodgett2020language,huang2023culturalbias,tao2024culturalbiasalignment,alkhamissi2024investigatingcultural,naous2024havingbeerprayermeasuring,sukiennik2025evaluationcultural}. We depart from this paradigm by operationalising culture as \emph{graded non-topical relevance}~\cite{manning2008ir,jarvelin2002cumulated,chapelle2009expected}. We propose the first \emph{leakage-free framework} using orthogonal supervision~\cite{li2025preferenceleakage} and intersection pools~\cite{voorhees2000trec,smucker2007comparison,carterette2012multipletesting,demsar2006statistical} to rigorously disentangle true ranking gains from circular evaluator preferences.

\definecolor{corpblue}{RGB}{235, 245, 255}    
\definecolor{darkblue}{RGB}{45, 85, 125}      
\definecolor{corpred}{RGB}{255, 240, 240}     
\definecolor{darkred}{RGB}{145, 45, 45}       
\definecolor{slate}{RGB}{240, 242, 245}       
\definecolor{charcoal}{RGB}{70, 75, 80}       

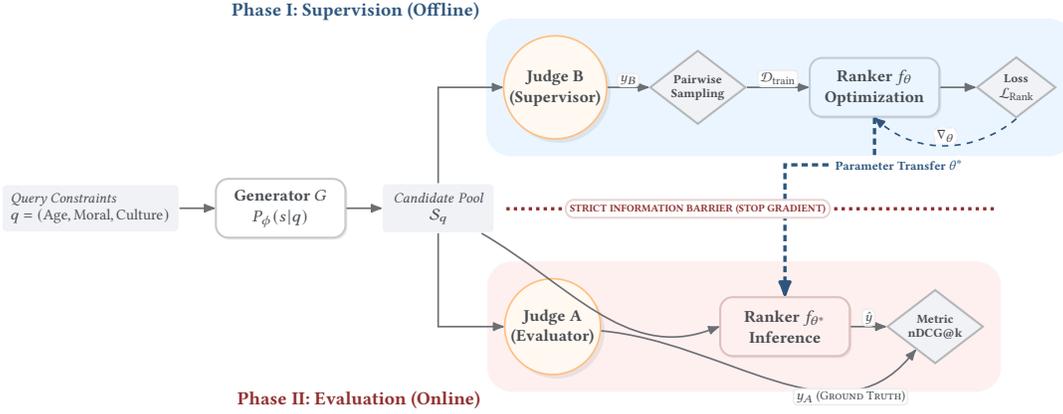
\begin{figure*}[t]
\centering
\resizebox{0.80\textwidth}{!}{%
\begin{tikzpicture}[
    node distance=1.2cm and 0.8cm, 
    font=\sffamily,
    >={LaTeX[round]},
    line width=1.0pt,
    model/.style={rectangle, rounded corners=6pt, minimum width=2.2cm, minimum height=1.0cm, draw=charcoal!30, fill=white, drop shadow={opacity=0.15, shadow xshift=1pt, shadow yshift=-1pt}, font=\bfseries\small, text=charcoal, align=center},
    judge/.style={circle, minimum size=1.6cm, draw=orange!50, thick, fill=orange!5, font=\bfseries\small, text=charcoal, align=center, drop shadow={opacity=0.15}, inner sep=1pt},
    op/.style={diamond, aspect=1.3, inner sep=1pt, draw=charcoal!40, fill=slate, font=\bfseries\scriptsize, text=charcoal, align=center},
    data/.style={rectangle, rounded corners=2pt, fill=slate, draw=none, font=\itshape\footnotesize, text=charcoal, inner sep=4pt},
    flow/.style={->, draw=charcoal!80, thick},
    grad/.style={->, draw=darkblue, dashed, thick},
    leakage/.style={draw=darkred, ultra thick, dotted},
    mathlabel/.style={fill=white, inner sep=1.0pt, rounded corners=2pt, text=charcoal, font=\scriptsize\scshape, draw=charcoal!10}
]
    \node[data, align=left] (query) {Query Constraints\\$q = ({\rm Age, Moral, Culture})$};
    \node[model, right=0.6cm of query] (generator) {Generator $G$\\$P_\phi(s|q)$};
    \node[data, right=0.6cm of generator, text width=1.6cm, align=center] (pool) {Candidate Pool\\$\mathcal{S}_q$};
    \draw[flow] (query) -- (generator);
    \draw[flow] (generator) -- (pool);
    
    \node[judge, above right=1.0cm and 0.4cm of pool] (judgeB) {Judge B\\(Supervisor)};
    \node[op, right=0.7cm of judgeB] (sampling) {Pairwise\\Sampling};
    \node[model, right=1.05cm of sampling, fill=corpblue, draw=darkblue!30] (ranker_train) {Ranker $f_\theta$\\Optimization};
    \node[op, right=0.6cm of ranker_train] (loss) {Loss\\$\mathcal{L}_{\rm Rank}$};
    
    \draw[flow] (pool) |- (judgeB);
    \draw[flow] (judgeB) -- node[mathlabel, above] {$y_B$} (sampling);
    \draw[flow] (sampling) -- node[mathlabel, above] {$\mathcal{D}_{\rm train}$} (ranker_train);
    \draw[flow] (ranker_train) -- (loss);
    \draw[grad] (loss.south) to[out=-140, in=-40] node[mathlabel, above] {$\nabla_\theta$} (ranker_train.south);
    
    \node[judge, below right=1.0cm and 0.4cm of pool] (judgeA) {Judge A\\(Evaluator)};
    \node[model, right=2.0cm of judgeA, fill=corpred, draw=darkred!30] (ranker_test) {Ranker $f_{\theta^*}$\\Inference};
    \node[op, right=0.6cm of ranker_test] (metric) {Metric\\nDCG@k};
    
    \draw[flow] (pool) |- (judgeA);
    \draw[flow] (judgeA) to[out=-5, in=-130] node[mathlabel, below, near end] {$y_A$ (Ground Truth)} (metric);
    \draw[flow] (ranker_test) -- node[mathlabel, above] {$\hat{y}$} (metric);
    \draw[flow] (pool) to[out=-30, in=200] (ranker_test.west);
    
    \draw[->, ultra thick, darkblue, densely dashed] (ranker_train.south) -- ++(0,-0.8) -| (ranker_test.north) node[pos=0.25, right, font=\bfseries\scriptsize, text=darkblue, fill=white, inner sep=2pt] {Parameter Transfer $\theta^*$};
    
    \coordinate (mid_v) at ($(judgeB.south)!0.5!(judgeA.north)$);
    \draw[leakage] ($(pool.east |- mid_v) + (0.2, 0)$) -- ($(metric.east |- mid_v) + (0.2, 0)$);
    \node[fill=white, text=darkred, font=\bfseries\tiny, draw=darkred!20, rounded corners, inner sep=2pt] at (sampling |- mid_v) {STRICT INFORMATION BARRIER (STOP GRADIENT)};
    
    \begin{scope}[on background layer]
        \node[fit=(judgeB) (loss) (ranker_train), fill=corpblue, rounded corners=12pt, inner sep=8pt] (zone_top) {};
        \node[above left, text=darkblue, font=\bfseries\normalsize, yshift=-4pt] at (zone_top.north west) {Phase I: Supervision (Offline)};
        
        \node[fit=(judgeA) (metric) (ranker_test), fill=corpred, rounded corners=12pt, inner sep=8pt] (zone_bot) {};
        \node[below left, text=darkred, font=\bfseries\normalsize, yshift=4pt] at (zone_bot.south west) {Phase II: Evaluation (Online)};
    \end{scope}
\end{tikzpicture}%
}
\vspace{-0.10in}
\caption{Our novel framework separates the optimisation loop (Top, Blue) from the inference/evaluation loop (Bottom, Red). The Supervision Judge ($J_B$) generates noisy proxies $y_B$ for gradient descent ($\nabla_\theta$). The Independent Evaluator ($J_A$) provides labels $y_A$ exclusively for metric computation. The only bridge between the worlds is the frozen parameter set $\theta^*$, preventing circular preference leakage. \(\hat{y} = f_{\theta^*}(s, q)\) is the framework's prediction.}
\label{fig:framework_sophisticated}
\end{figure*}
\vspace{-3mm}
\section{Our Novel Framework}
\label{sec:methodology}
We propose a novel framework, depicted in Figure~\ref{fig:framework_sophisticated}, for culturally grounded story selection that treats cultural relevance as a \emph{within-query graded ranking} problem. Our approach addresses the ``label leakage'' and circular self-preference plaguing current GenIR evaluation by enforcing strict role separation between ranking supervision and independent assessment.

\emph{\textbf{Selection vs.\ Model-Based Retrieval:}} While some GenIR directions (e.g., DSI) internalise a corpus into model parameters~\cite{tay2022transformermemory}, our work targets the \emph{post-generation selection bottleneck}: ranking multiple valid candidates produced under identical constraints, common to RAG and creative generation pipelines~\cite{lewis2020retrievalaugmented}.

\emph{\textbf{Culture as Graded Non-Topical Relevance:}} Unlike prior work that treats culture as a binary bias axis, we operationalise \emph{cultural relevance} as a graded, non-topical relevance signal---consistent with multidimensional relevance criteria in classical IR---enabling within-query comparisons under fixed cutoffs using standard graded metrics~\cite{jarvelin2002cumulated,manning2008ir}.

\subsection{Formal Problem Definition}
To ensure rigorous evaluation, we ground our generation constraints in established sociological and educational constructs. We define \textbf{Target Age} as a constraint on cognitive accessibility and linguistic complexity, scaling from concrete operational stages to abstract reasoning. \textbf{Moral Theme} operationalises the narrative's pro-social intent, requiring the resolution of conflict through specific virtues (e.g., honesty, patience) rather than generic happy endings. We define \textbf{Cultural Context} not merely as a geographic label, but as a ``thick'' normative setting comprised of specific rituals, etiquette, and social dynamics that govern how the moral is enacted. Together, these three dimensions constitute the normative relevance criteria against which the generator's alignment is measured.

We operationalise the query $q$ as a triple: Target Age ($c_{\text{age}}$) constrains complexity; Moral Theme ($c_{\text{moral}}$) defines the plot resolution; and Cultural Context ($c_{\text{cult}}$) provides the normative setting (e.g., Arab, Latin American), requiring grounding in customs beyond surface-level keywords.

While we treat Target Age and Moral Theme as explicit generation constraints that define the content boundary, we view Cultural Context as the primary source of normative variance. Thus, our evaluation focuses on distinguishing between surface-level thematic compliance (e.g., getting the moral right) and deep normative enactment (e.g., behaving appropriately within that moral context).

To guarantee valid outputs, we employ a rigid generation template that enforces a complete narrative arc, a strictly delimited moral resolution, and the exclusion of meta-commentary to ensure consistent parsing. The full prompt structure, which instantiates constraints such as $q=(8, \text{Honesty}, \text{Arab})$ into specific instruction slots, is provided in the released dataset artefacts.

These three constraints constitute a minimal yet comprehensive framework that captures the essential cognitive, thematic, and normative dimensions of educational storytelling. Limiting the query to this triad isolates the core challenge of normative alignment without conflating the evaluation with purely stylistic preferences.

\emph{\textbf{Mathematical Formulation:}} Formally, let $\mathcal{Q}$ be the space of structured queries, where each $q \in \mathcal{Q}$ is defined by the tuple $q = (c_{\text{age}}, c_{\text{moral}}, c_{\text{cult}})$, where $c_{\text{age}} \in \{4, \dots, 10\}$ is the integer age constraint, $c_{\text{moral}}$ is a categorical variable from the set of morals $\mathcal{M}$, and $c_{\text{cult}}$ is a categorical variable from the set of cultures $\mathcal{C}$. Let $\mathcal{S}$ be the infinite space of possible narratives. We assume a stochastic generator $G$ parameterised by $\phi$ that defines a conditional probability distribution $P_\phi(s | q)$ over $\mathcal{S}$.

For a given query $q$, we sample a candidate set $\mathcal{S}_q = \{s_1, \dots, s_{n_q}\}$ where $s_i \sim P_\phi(\cdot | q)$, and \(n_q\)
denotes the size of the candidate pool $|\mathcal{S}_q|$. We assume the existence of a latent target scoring function (the ``ideal evaluator'') $R^*: \mathcal{S} \times \mathcal{Q} \rightarrow \mathcal{Y}$, where $\mathcal{Y} = \{1, \dots, 5\}$ is the ordinal relevance space. In practice, $R^*$ is unobserved. We approximate it with rubric-based LLM judges under a strict two-judge framework, where Judge~B provides supervision for learning, and Judge~A is used only for final, leakage-free evaluation. 
Our goal is to learn a ranking function $f_\theta: \mathcal{S} \times \mathcal{Q} \rightarrow \mathbb{R}$ such that for any pair $s_i, s_j \in \mathcal{S}_q$: \(f_\theta(s_i, q) > f_\theta(s_j, q) \iff R^*(s_i, q) > R^*(s_j, q)\).

\emph{\textbf{Two-Judge Leakage-Free Framework:}} A central requirement is strict separation between ranking supervision and evaluation labels to prevent inflated estimates due to circular evaluation~\cite{voorhees2000trec,smucker2007comparison,blackwell2025reproduciblellm,laskar2024systematicllmeval}. We employ two rubric-guided judges~\cite{Hashemi2024llmrubric,zheng2023mtbenchjudge,li2024llmsasjudgescomprehensive} with non-overlapping roles.

For $\mathcal{Y} = \{1, \dots, 5\}$ (i.e., the space of ordinal relevance grades) and a generated story $s$ given query $q$: \textbf{Judge~B (Supervision):} Assigns a score $y_B(s, q) \in \mathcal{Y}$ used \emph{only} for ranking supervision. \textbf{Judge~A (Evaluation):} Assigns an independent label $y_A(s, q)$ $\in \mathcal{Y}$ used \emph{only} for final assessment.

Judge~A is never used for training, early stopping, hyperparameter tuning, prompt tuning, threshold selection, or any form of model selection. Formally, we enforce strict role separation: $y_B$ is used only to train/tune $f_\theta$, while all reported metrics are computed only from $y_A$ on held-out queries. This mitigates preference leakage and self-preference bias, where evaluators disproportionately favour outputs from their own model family~\cite{li2025preferenceleakage,koo2024benchmarkingcognitive,chen2024humansllmsjudgestudy}. We enforce this by instantiating Judge~A and Judge~B using different model families (e.g., Yi-34 vs. Llama-3). We employ a hierarchical rubric where Score 1 indicates irrelevant content, Score 3 indicates generic thematic alignment without cultural markers, and Score 5 requires deep integration of specific cultural customs and worldview.

\emph{\textbf{Reasoning-Aware Judge Prompting:}}
To encourage judges to evaluate the narrative substance rather than surface patterns, we employ a Chain-of-Thought (CoT) style prompting \cite{barrot2026generative} setup that asks the judge to consider (internally) the cultural customs and moral integration given the constraint triple $q$ and the rubric, and then output only a final score. Particularly, both judges receive the story text, $q$, and the rubric, and are instructed to reflect on cultural grounding and moral realisation before assigning a single integer score. To preserve strict output constraints and prevent extraneous text, we require JSON-only outputs and do not request or store intermediate reasoning. This reasoning-aware setup reduces the noise often seen in direct-scoring approaches~\cite{wei2022chain}.

\emph{\textbf{Unified Candidate Pool (Intersection):}} To ensure all methods are compared on identical candidates and to avoid missing-label confounds, we define the evaluation pool per query as the intersection of valid Judge~A and Judge~B annotations:
\[
\mathcal{S}_q^{\cap} = \{ s \in \mathcal{S}_q : y_A(s,q)\ \text{valid} \ \wedge\  y_B(s,q)\ \text{valid} \}.
\]
All methods rank candidates within $\mathcal{S}_q^{\cap}$. We compute metrics only for queries with $|\mathcal{S}_q^{\cap}|\ge k$ at cutoff $k{=}5$~\cite{manning2008ir,jarvelin2002cumulated}. 
We split data at the query level: all candidates for the same constraint tuple $q$ remain in a single partition (train/dev/test) to prevent cross-query leakage.

\emph{\textbf{Dense Bi-Encoder:}} We include a dense bi-encoder baseline based on BGE-M3~\cite{Xiao2023bge}. For each query $q$, we form a deterministic text serialisation $T(q)$ from the constraint triple (age, moral, culture), and encode $T(q)$ and each candidate story $s \in \mathcal{S}_q^{\cap}$ independently. Candidates are ranked by cosine similarity in embedding space, providing an efficient semantic baseline under the same fixed candidate pools used throughout.

We additionally report a teacher--student distilled bi-encoder (BGE-M3 Distilled). This model is implemented in SentenceTransformers and trained offline on the NGR-33k intersection training split using a cosine-similarity distillation objective, where the student bi-encoder is optimised to match a teacher-produced similarity signal derived from Judge~B–supervised cross-encoder training. At inference, the distilled student encodes $T(q)$ and $s$ independently and ranks by cosine similarity, yielding a lightweight dense model that transfers the teacher’s supervision while preserving bi-encoder efficiency and full locality/reproducibility.

To distil the normative reasoning of the Judge-B-supervised Cross-Encoder (Teacher) into the efficient BGE-M3 Bi-Encoder (Student), we minimise the Mean Squared Error (MSE) between the teacher's interaction-based relevance logits and the student's cosine similarity scores. Formally, for a query $q$ and candidate $s$, the distillation loss $\mathcal{L}_{\text{distill}}$ is defined as:
$$\mathcal{L}_{\text{distill}} = \frac{1}{|\mathcal{B}|} \sum_{(q,s) \in \mathcal{B}} \left( \sigma(f_{\text{Teacher}}(q, s)) - \cos(\mathbf{e}_q, \mathbf{e}_s) \right)^2,$$
where $f_{\text{Teacher}}$ is the logit output of the Cross-Encoder, $\sigma$ is the sigmoid activation, and $\mathbf{e}_q, \mathbf{e}_s$ are the embeddings produced by the student. This pointwise regression allows the student to approximate the teacher's latent normative manifold, while the bi-encoder architecture acts as a bottleneck to filter judge-specific noise.

\emph{\textbf{Lightweight Neural Rankers over Frozen Embeddings:}} For the supervised neural rankers (pointwise and pairwise), we freeze a dense encoder $E:\mathcal{X}\rightarrow\mathbb{R}^d$ (BGE-M3) and learn a small scoring function on top of its representations. We define a deterministic serialisation function $T: \mathcal{Q} \rightarrow \mathcal{V}^*$ that maps the constraint tuple to a natural language string (e.g., $q \mapsto \text{``Age: 8, Moral: Honesty...''}$). We map the query $q$ and candidate story $s$ to a joint latent representation $\mathbf{h} \in \mathbb{R}^{2d}$:
\[
\mathbf{h}(s, q) = [E(T(q)) \ ; \ E(s)],
\]
where $T(q)$ is the textual serialisation of the constraint tuple and $[\cdot ; \cdot]$ denotes vector concatenation.
The scoring function $f_\theta$ is parameterised as a Multi-Layer Perceptron (MLP) mapping the latent state to a scalar relevance score:
\[
f_\theta(s, q) = \mathbf{w}_2^\top \text{ReLU}(\mathbf{W}_1 \mathbf{h}(s, q) + \mathbf{b}_1) + b_2,
\]
where $\theta = \{\mathbf{W}_1, \mathbf{b}_1, \mathbf{w}_2, b_2\}$ represents the learnable parameters.

\emph{\textbf{Objective 1: Pointwise Risk Minimisation:}}
We treat the rubric scores from Judge~B, denoted as $y_B(s, q)$, as noisy proxies for $R^*$. We frame the learning task as minimising the empirical risk under the Mean Squared Error (MSE) criterion against the discrete integer scores $y_B$. The objective $\mathcal{L}_{\text{MSE}}(\theta)$ is given by:
\[
\mathcal{L}_{\text{MSE}}(\theta) = \mathbb{E}_{(s, q) \sim \mathcal{D}_{\text{train}}} \left[ \left\| f_\theta(s, q) - \frac{y_B(s, q)}{y_{\max}} \right\|^2_2 \right],
\]
where $y_{\max}=5$ is the scaling factor to normalise targets to $[0, 1]$.

\emph{\textbf{Objective 2: Pairwise Probabilistic Ranking:}}
We construct a set of preferences $\mathcal{P}_q = \{(i, j) \mid s_i, s_j \in \mathcal{S}_q, y_B(s_i) > y_B(s_j)\}$. We model the probability of $s_i \succ s_j$ using a logistic function of the score differences:
{\small \[
P(s_i \!\succ\! s_j \mid \theta)
= \sigma\!\big(f_\theta(s_i,q)-f_\theta(s_j,q)\big)
= \frac{1}{1 \!+\! \exp\!\left(-(f_\theta(s_i,q)\!-\!f_\theta(s_j,q))\right)}.
\]}  \hspace{-0.55em}
We optimise $\theta$ by minimising the cross-entropy loss between the Judge~B preferences and the model's induced probability distribution:
\[
\mathcal{L}_{\text{RankNet}}(\theta) = - \sum_{q \in \mathcal{Q}} \sum_{(i, j) \in \mathcal{P}_q} \log P(s_i \succ s_j | \theta).
\]

\section{Experiments and Results}
\label{sec:experimental_setup}
We study GenIR in a \emph{generate-then-rank} setting, where multiple candidates are generated under identical constraints and the bottleneck becomes \emph{selection} rather than generation~\cite{Alaofi2024generativeinformation,lewis2020retrievalaugmented,bonifacio2022inpars,manning2008ir}.

\subsection{Data and Query Construction}
\label{sec:data}

\emph{\textbf{Story Corpus and Structured Queries:}} We generated 35{,}000 children's stories (Table~\ref{tab:data_distribution_gemma}) under explicit constraints for age, moral theme, and cultural context. After pre-processing and schema validation, we retained 33{,}052 stories for downstream judging and ranking. Stories are generated offline using Gemma-2-9B-IT~\cite{gemma2report2024} with a fixed prompt template and stochastic decoding.

We define the query space $\mathcal{Q}$ as the Cartesian product of 7 ages, 7 morals, and 7 cultures, yielding $|\mathcal{Q}| = 343$ distinct constraint tuples. We targeted a corpus size of $N=33{,}052$ to enforce dense candidate distributions ($\mu \approx 96$ per query), which are mathematically necessary to minimise estimator variance in top-$k$ rank metrics, specifically at our target $k{=}5$. This high density ensures sufficient support for constructing stable pairwise preference graphs, enabling the learning of fine-grained discriminative boundaries often impossible in sparse collections. Unlike standard benchmarks where top-$k$ stability is brittle, our design guarantees a robust signal within every query partition. This structure is critical for reliable variance estimation and reproducible query-level evaluation~\cite{voorhees2000trec,smucker2007comparison}.

We deterministically derive a query identifier to group stories into within-query candidate sets following standard IR practice of evaluating ranked lists \emph{within} a query context.

\begin{table}[t]
\centering
\caption{Dataset ($N=33{,}052$)---balanced distribution across dimensions.}
\label{tab:data_distribution_gemma}
\vspace{-0.15in} 
\resizebox{\columnwidth}{!}{
\renewcommand{\arraystretch}{1.1} 
\setlength{\tabcolsep}{3pt}
\begin{tabular}{l l}
\toprule
\textbf{Dimension} & \textbf{Distribution (Value: Count)} \\
\midrule
\rowcolor{blue!5}
\textbf{Age Group} & 4 (4,713); 5 (4,895); 6 (4,568); 7 (4,773); 8 (4,412); 9 (4,947); 10 (4,744) \\
\rowcolor{orange!5}
\textbf{Culture} & African (4,831); Arab (4,729); E. Asian (4,628); European (4,653); Global/Neut. (4,391); Lat. Am. (5,102); S. Asian (4,718) \\
\rowcolor{green!5}
\textbf{Moral Theme} & Bravery (4,817); Cooperation (4,828); Empathy (4,818); Honesty (4,529); Kindness (4,755); Patience (4,602); Respect (4,703) \\
\bottomrule
\end{tabular}
}
\vspace{-5mm} 
\end{table}

The intersection filtering removes missing/invalid judge outputs and ensures that every method ranks \emph{the same candidates} per query, which is critical for fair within-query comparisons under judge-based evaluation~\cite{voorhees2000trec,smucker2007comparison}. On the processed NGR-33k dataset ($N=33{,}052$), the unified intersection pool $(y_A \cap y_B)$ retains 17{,}965 stories across 343 queries. Across eligible queries, intersection-pool sizes satisfy $\min/\mathrm{median}/\mathrm{mean}/\max = 10/52/52.38/106$, and all reported results use cutoff $k{=}5$ and evaluate only queries with $|\mathcal{S}_q^{\cap}|\ge k$.

We construct 10 query-level splits (10 seeds) so that each \texttt{query\_id} and its full candidate set appear in exactly one of train/dev/test per seed. In the main setting, each seed contains 239/34/70 train/dev/test queries, i.e., 70 held-out test queries per seed. We compute metrics per query and macro-average across the 70 test queries to obtain a seed-level score, and then report mean$\pm$std over the 10 seeds.
Using 10 query-level splits follows standard robustness evaluation practice in IR, enabling variance estimation across random partitions without overfitting to a single test split~\cite{voorhees2000trec,smucker2007comparison}.

To provide an out-of-domain check, we also evaluate on the publicly available SSGEN~\cite{li2024ssgen} and Moral Stories datasets \cite{emelin2021moral}. We uniformly sample $N{=}500$ instances with a fixed random seed (42) and score the same subset independently with Judge~A and Judge~B. We report each judge's failure rate (null outputs) and inter-judge agreement on the intersection of instances with valid scores.

\emph{\textbf{Judges and Prompts:}} \textbf{Judge~A} (Yi-1.5-9B-Chat~\cite{01ai2024yi15modelcard}, evaluation-only) and Judge~B (LLaMA-3.1-8B-Instruct~\cite{meta2024llama3herd}, (supervision-only) are used to enforce strict role separation~\cite{koo2024benchmarkingcognitive,chen2024humansllmsjudgestudy,li2025preferenceleakage,li2024llmsasjudgescomprehensive}. Both use a hierarchical 1--5 rubric scoring only cultural relevance via strict JSON. Judge~A includes additional anti-default guidance to prevent score compression.

\emph{\textbf{Decoding, Format Constraints, and Validity:}} All judges use deterministic decoding (temperature $=0$, top-$p=1.0$) and output a single integer score in $\{1,\dots,5\}$ via strict JSON formatting. A label is valid only if (i) parsing succeeds and (ii) the parsed score lies in range; otherwise, the annotation is treated as missing. Missing/invalid outputs are handled solely by intersection filtering into $\mathcal{S}_q^{\cap}$ to keep candidate pools identical across methods.

\emph{\textbf{Judge Validity and Intersection Retention:}} On the processed NGR-33k dataset ($N=33{,}052$), Judge~A produces 8{,}524 null or invalid outputs (25.79\%), while Judge~B produces 10{,}912 null or invalid outputs (33.01\%) under strict JSON parsing and in-range checks. Judge B is instantiated using LLaMA 3.1 8B Instruct, which exhibits conservative scoring behaviour and produces null outputs more frequently under strict JSON parsing and in-range checks. Approximately 33\% of generated stories receive null supervision signals, reflecting selective supervision rather than annotation noise. All ranking models are trained and evaluated on the non-null intersection, ensuring consistent and leakage-free comparison. To ensure fair and leakage-free comparison, all methods are evaluated on the unified intersection pool $(y_A \cap y_B)$, which retains 17{,}965 stories across all 343 queries. All queries remain eligible at the cutoff $k{=}5$.

\begin{table}[t]
\centering
\caption{Raw generated stories and filter for valid judge outputs. \colorbox{green!5}{\textbf{Intersection Pool}} is used for evaluation.}
\label{tab:pool_stats_q70}
\vspace{-0.10in}
\resizebox{0.90\columnwidth}{!}{
\setlength{\tabcolsep}{8pt}

\begin{tabular}{lrrrr}
\toprule
\textbf{Candidate Pool Step} & \textbf{Min} & \textbf{Med} & \textbf{Mean} & \textbf{Max} \\
\midrule
1. Raw Generated ($|\mathcal{S}_q|$) & 33 & 95 & 95.7 & 161 \\
\addlinespace[0.1cm]

2. Judge A Valid ($|\mathcal{S}_q^A|$) & 20 & 69 & 70.7 & 125 \\
3. Judge B Valid ($|\mathcal{S}_q^B|$) & 12 & 63 & 64.2 & 135 \\
\addlinespace[0.1cm]

\rowcolor{green!5}
\textbf{4. Intersection} ($|\mathcal{S}_q^{\cap}|$) & \textbf{10} & \textbf{51} & \textbf{51.7} & \textbf{106} \\

\midrule
\multicolumn{5}{c}{\textit{Eligibility Check: 100\% of queries satisfy $|\mathcal{S}_q^{\cap}| \ge 5$}} \\
\bottomrule
\end{tabular}
}
\vspace{-1mm}
\end{table}

No candidate truncation is applied: all methods rank the full intersection pool $\mathcal{S}_q^{\cap}$ per query. This avoids introducing method-specific biases arising from heuristic pre-filtering.

\begin{table}[t]
\centering
\caption{Ranking performance on NGR-33k ($k{=}5$). We compare \textit{Unsupervised Baselines}, \textit{Teacher--Student Distillation}, and \textit{Judge-B supervised} methods. \textbf{Bold} indicates the best performing model. Note that the Distilled Student outperforms the Teacher (Cross-Encoder).}
\label{tab:main_results}
\vspace{-0.10in}

\renewcommand{\arraystretch}{1.2} 
\resizebox{\columnwidth}{!}{
\setlength{\tabcolsep}{6pt}      
\begin{tabular}{lcccc}
\toprule
 & \multicolumn{2}{c}{\textbf{Effectiveness}} & \multicolumn{2}{c}{\textbf{Judge Alignment}} \\
\cmidrule(lr){2-3} \cmidrule(lr){4-5}
\textbf{Method} & \textbf{nDCG@5} & \textbf{ERR@5} & $\boldsymbol{\tau_b}$ & $\boldsymbol{\rho}$ \\
\midrule

\multicolumn{5}{l}{\textit{\textbf{Unsupervised Baselines}}} \\
\hspace{3mm} Random 
    & $0.571{\scriptsize\pm0.03}$ 
    & $0.817{\scriptsize\pm0.02}$ 
    & -- & -- \\
\hspace{3mm} BM25 
    & $0.615{\scriptsize\pm0.04}$ 
    & \textbf{0.851}${\scriptsize\pm0.01}$ 
    & $0.134$ 
    & $0.206$ \\
\hspace{3mm} DPH 
    & $0.616{\scriptsize\pm0.04}$ 
    & $0.501{\scriptsize\pm0.02}$ 
    & $0.071{\scriptsize\pm0.02}$ 
    & $0.086{\scriptsize\pm0.02}$ \\
\hspace{3mm} Dirichlet LM 
    & $0.612{\scriptsize\pm0.04}$ 
    & $0.502{\scriptsize\pm0.02}$ 
    & $0.064{\scriptsize\pm0.02}$ 
    & $0.078{\scriptsize\pm0.02}$ \\

\addlinespace[0.6em]

\multicolumn{5}{l}{\textit{\textbf{Judge-B Supervised Ranking (Direct)}}} \\
\hspace{3mm} Neural Pointwise
    & $0.574{\scriptsize\pm0.03}$
    & $0.489{\scriptsize\pm0.02}$
    & $0.016{\scriptsize\pm0.02}$
    & $0.020{\scriptsize\pm0.02}$ \\
\hspace{3mm} Neural Pairwise 
    & $0.568{\scriptsize\pm0.04}$ 
    & $0.799{\scriptsize\pm0.01}$ 
    & $0.089{\scriptsize\pm0.01}$ 
    & $0.116{\scriptsize\pm0.01}$ \\
\hspace{3mm} Listwise (ListMLE)
    & $0.710{\scriptsize\pm0.02}$ 
    & $0.788{\scriptsize\pm0.02}$ 
    & $0.053{\scriptsize\pm0.02}$ 
    & $0.066{\scriptsize\pm0.02}$ \\

\addlinespace[0.6em]

\multicolumn{5}{l}{\textit{\textbf{Teacher--Student Distillation}}} \\
\rowcolor{blue!5}
\hspace{3mm} \textbf{BGE-M3 (Student)} 
    & $\mathbf{0.771}{\scriptsize\pm0.02}$ 
    & $0.614{\scriptsize\pm0.02}$ 
    & $\mathbf{0.313}{\scriptsize\pm0.03}$ 
    & $\mathbf{0.377}{\scriptsize\pm0.03}$ \\

\addlinespace[0.6em]

\multicolumn{5}{l}{\textit{\textbf{Judge Signals (Reference Only)}}} \\
\hspace{3mm} B-Score (Oracle $y_B$) 
    & $0.689{\scriptsize\pm0.02}$ 
    & $0.837{\scriptsize\pm0.02}$ 
    & $0.297{\scriptsize\pm0.04}$ 
    & $0.326{\scriptsize\pm0.04}$ \\
\hspace{3mm} Cross-Encoder (Teacher) 
    & \underline{$0.577{\scriptsize\pm0.03}$} 
    & \underline{$0.479{\scriptsize\pm0.01}$} 
    & \underline{$0.092{\scriptsize\pm0.01}$} 
    & \underline{$0.121{\scriptsize\pm0.01}$} \\

\bottomrule
\end{tabular}
}
\vspace{-0.4em}
\end{table}

\emph{\textbf{Story Generation and Query Grouping:}} Stories are generated offline under structured constraints $q$ using a fixed prompt template. Each story is stored with a unique \texttt{story\_id} (Table~\ref{tab:pool_stats_q70}), and we deterministically derive a \texttt{query\_id} from the constraint triple to group stories into within-query candidate pools $\mathcal{S}_q$.

\begin{table}[t]
\centering
\caption{Late-Interaction Performance. ColBERTv2 offers a middle ground between bi-encoders and cross-encoders.}
\label{tab:colbert_results}
\vspace{-0.1in}

\renewcommand{\arraystretch}{1.2}

\resizebox{\columnwidth}{!}{
\setlength{\tabcolsep}{8pt}
\begin{tabular}{lcccc}
\toprule
\textbf{Interaction Mechanism} & \textbf{nDCG} & \textbf{ERR} & $\boldsymbol{\tau_b}$ & $\boldsymbol{\rho}$ \\
\midrule

\rowcolor{blue!5}
ColBERTv2 (Late-Interaction)
& $0.635${\scriptsize$\pm.06$} 
& $0.707${\scriptsize$\pm.04$} 
& $0.209${\scriptsize$\pm.04$} 
& $0.250${\scriptsize$\pm.04$} \\

\bottomrule
\end{tabular}
}
\vspace{-0.5em}
\end{table}

\emph{\textbf{Query-Level Splits:}} We evaluate using query-level splits (10 seeds, 70 test queries per seed). Candidates from the same query never span across train/test sets. Query-level splitting is essential in this setting, as splitting at the story level would induce information leakage by allowing candidates generated under identical constraints to appear across train and test partitions. All hyperparameters are selected using development queries under Judge~B supervision, ensuring zero leakage of Judge~A signals into the model pipeline.

\emph{\textbf{Aggregation:}} For each seed, we compute metrics per query and macro-average over eligible test queries to obtain a seed-level score, and then report mean$\pm$std over the 10 seeds. For rank correlations ($\tau_b$, $\rho$), we compute correlation \emph{per query} and macro-average across queries; correlations may be undefined for degenerate queries (e.g., all Judge~A grades tied), in which case the query is excluded from correlation aggregation and the count of defined queries is tracked.

\begin{table}[t]
\centering
\small
\caption{Paired sign-flip permutation tests (50k permutations; $n{=}700$ aligned query$\times$seed rows from the same stack as the main table). We report mean differences ($\Delta=\mathbb{E}[A-B]$) and two-sided $p$-values. \textbf{Bold} indicates statistical significance ($p<0.05$). $^{*}$ denotes a significant decrease.}
\label{tab:q70_significance_bge}
\vspace{-0.10in}
\resizebox{\columnwidth}{!}{
\setlength{\tabcolsep}{5pt}
\begin{tabular}{lcccc}
\toprule
\textbf{Comparison (A vs.\ B)} 
& $\Delta$\textbf{nDCG@5} & \textbf{$p$-val} 
& $\Delta$\textbf{ERR@5} & \textbf{$p$-val} \\
\midrule

\multicolumn{5}{l}{\textit{\textbf{1. Improvements over Random}}} \\

BM25 vs.\ Random 
& \textbf{+0.044} & \textbf{$<10^{-4}$} 
& \textbf{+0.034} & \textbf{$<10^{-4}$} \\

B-Score ($y_B$) vs.\ Random 
& \textbf{+0.118} & \textbf{$<10^{-4}$} 
& \textbf{+0.020} & \textbf{0.019} \\

Neural Pairwise vs.\ Random 
& -0.003 & 0.664 
& \textbf{-0.018}$^{*}$ & \textbf{0.035} \\

Cross-Encoder ($y_B$) vs.\ Random 
& +0.007 & 0.322 
& \textbf{-0.338}$^{*}$ & \textbf{$<10^{-4}$} \\

BGE-M3 (Distilled) vs.\ Random
& \textbf{+0.384} & \textbf{$<10^{-4}$}
& \textbf{+0.119} & \textbf{$<10^{-4}$} \\

\midrule
\multicolumn{5}{l}{\textit{\textbf{2. Head-to-Head Comparisons}}} \\

B-Score ($y_B$) vs.\ BM25 
& \textbf{+0.074} & \textbf{$<10^{-4}$} 
& \textbf{-0.014}$^{*}$ & \textbf{0.038} \\

Cross-Encoder ($y_B$) vs.\ BM25 
& \textbf{-0.038}$^{*}$ & \textbf{$<10^{-4}$} 
& \textbf{-0.372}$^{*}$ & \textbf{$<10^{-4}$} \\

Cross-Encoder ($y_B$) vs.\ B-Score ($y_B$) 
& \textbf{-0.112}$^{*}$ & \textbf{$<10^{-4}$} 
& \textbf{-0.358}$^{*}$ & \textbf{$<10^{-4}$} \\

BGE-M3 (Distilled) vs.\ BM25
& \textbf{+0.340} & \textbf{$<10^{-4}$}
& \textbf{+0.085} & \textbf{$<10^{-4}$} \\

BGE-M3 (Distilled) vs.\ B-Score ($y_B$)
& \textbf{+0.266} & \textbf{$<10^{-4}$}
& \textbf{+0.099} & \textbf{$<10^{-4}$} \\

\bottomrule
\end{tabular}
}
\vspace{-0.3em}
\end{table}

\begin{table}[t]
\centering
\caption{Sensitivity to pool density ($|\mathcal{S}_q^{\cap}|$, nDCG@5). \textbf{BGE-M3 (Distilled)} yields robust gains across all regimes, whereas direct neural rankers fail to outperform baselines.}
\label{tab:pool_sensitivity_main}
\vspace{-0.10in}
\resizebox{0.95\columnwidth}{!}{
\renewcommand{\arraystretch}{1.15}
\setlength{\tabcolsep}{8pt}

\begin{tabular}{lccc}
\toprule
& \multicolumn{3}{c}{\textbf{Intersection Pool Density}} \\
\cmidrule(lr){2-4}
\textbf{Method} & \textbf{Small} ($\le 35$) & \textbf{Medium} ($36\text{--}50$) & \textbf{Large} ($>50$) \\
\midrule

\multicolumn{4}{l}{\textit{Lexical \& Random Baselines}} \\
Random & $0.592${\scriptsize$\pm.04$} & $0.555${\scriptsize$\pm.04$} & $0.567${\scriptsize$\pm.03$} \\
BM25   & $0.607${\scriptsize$\pm.06$} & $0.607${\scriptsize$\pm.05$} & $0.623${\scriptsize$\pm.06$} \\
\addlinespace

\multicolumn{4}{l}{\textit{Dense Retrieval Baseline}} \\
\rowcolor{blue!5}
\textbf{BGE-M3 (Distilled)} & $\mathbf{0.755}${\scriptsize$\pm.07$} & $\mathbf{0.744}${\scriptsize$\pm.07$} & $\mathbf{0.776}${\scriptsize$\pm.04$} \\
\addlinespace

\multicolumn{4}{l}{\textit{Neural Distillation ($y_B$ Supervision)}} \\
N. Pointwise & $0.588${\scriptsize$\pm.04$} & $0.559${\scriptsize$\pm.04$} & $0.576${\scriptsize$\pm.04$} \\
N. Pairwise  & $0.582${\scriptsize$\pm.04$} & $0.552${\scriptsize$\pm.06$} & $0.570${\scriptsize$\pm.04$} \\

\bottomrule
\end{tabular}
}
\vspace{-0.5em}
\end{table}

\subsection{Comparative Models}
We evaluate three classes of ranking strategies. All methods rank candidates within the strictly defined intersection pool $\mathcal{S}_q^{\cap}$ and are evaluated exclusively against Judge~A labels ($y_A$).

\emph{\textbf{1. Unsupervised Baselines:}} To measure the difficulty of the task without normative supervision, we employ:
\textbf{Random:} A uniform random permutation (lower bound). \textbf{Lexical Rankers:} We evaluate  \textbf{BM25}~\cite{robertson2009probabilistic}, \textbf{DPH} (Divergence-from-Randomness)~\cite{amati2002dfr}, and \textbf{Dirichlet LM} ($\mu{=}2000$)~\cite{zhai2004study}. All lexical models score a pseudo-query derived from the constraint string $(\text{age}, \text{moral}, \text{culture})$.

\begin{table}[t]
\centering
\caption{Pairwise ranker performance vs. training data fraction. \colorbox{green!5}{\textbf{Full Supervision}} is required for significant gains.}
\label{tab:q70_ablation_frac}
\vspace{-0.05in}
\resizebox{0.75\columnwidth}{!}{
\setlength{\tabcolsep}{8pt}

\begin{tabular}{lcc}
\toprule
\textbf{Training Data \%} & \textbf{nDCG@5} & \textbf{ERR@5} \\
\midrule
25\% (Sparse) & $.574${\scriptsize$\pm.03$} & $.496${\scriptsize$\pm.02$} \\
50\% & $.573${\scriptsize$\pm.03$} & $.492${\scriptsize$\pm.01$} \\
75\% & $.574${\scriptsize$\pm.03$} & $.490${\scriptsize$\pm.01$} \\

\rowcolor{green!5}
\textbf{100\% (Full)} & $\mathbf{.585}${\scriptsize$\pm.03$} & $\mathbf{.501}${\scriptsize$\pm.01$} \\
\bottomrule
\end{tabular}
}
\vspace{-5mm}
\end{table}

\begin{table*}[t]
\centering
\caption{We contrast top-ranked stories for three queries. \textbf{Bold text} highlights specific cultural grounding cues. \colorbox{corpblue}{\textbf{B-Score (Ours)}} prioritises narratives where the moral is enacted through specific cultural practices (e.g., \textit{Rangoli}, \textit{Sharing Bread}), whereas the baseline selects generic thematic matches.}
\label{tab:qual_comparison}
\vspace{-0.15in} 
\resizebox{\textwidth}{!}{
\renewcommand{\arraystretch}{0.9} 
\setlength{\tabcolsep}{3pt}       

\begin{tabular}{p{0.14\textwidth} p{0.42\textwidth} >{\columncolor{corpblue!40}}p{0.42\textwidth}}
\toprule
\textbf{Query Constraints} & \textbf{Dense Baseline (BGE-M3)} & \textbf{B-Score (Judge-Driven)} \\
\midrule

\footnotesize 
\textbf{Age:} 4 \newline
\textbf{Moral:} Patience \newline
\textbf{Culture:} S. Asian &
\footnotesize
\emph{``Maya loved \textbf{mangoes} and asked every day when they would be ready. Her grandmother told her to wait. When the mangoes finally ripened, Maya learned that good things take time.''} \newline
{\scriptsize\color{charcoal}\textit{$\rightarrow$ Generic trope (fruit ripening).}} &
\footnotesize
\emph{``Maya loved to play \textbf{Holi}. She wanted to start right away, but her mother asked her to help prepare the \textbf{rangoli} first. As Maya carefully worked, she learned that waiting and preparing together made the celebration more joyful.''} \newline
{\scriptsize\color{darkblue}\textbf{$\rightarrow$ Grounded in specific festival practices.}} \\
\addlinespace[0.25em] 

\footnotesize
\textbf{Age:} 6 \newline
\textbf{Moral:} Kindness \newline
\textbf{Culture:} African &
\footnotesize
\emph{``Ayo was kind to his classmates at school. He helped them and felt happy because \textbf{being kind is important}.''} \newline
{\scriptsize\color{charcoal}\textit{$\rightarrow$ Didactic / Tell-don't-show.}} &
\footnotesize
\emph{``In the village, Ayo noticed his friend had no lunch. He \textbf{broke his own bread} in half and shared it. Soon, other children joined, and everyone \textbf{ate together under the big tree}.''} \newline
{\scriptsize\color{darkblue}\textbf{$\rightarrow$ Enacted via communal sharing customs.}} \\
\addlinespace[0.25em] 

\footnotesize
\textbf{Age:} 6 \newline
\textbf{Moral:} Kindness \newline
\textbf{Culture:} Global &
\footnotesize
\emph{``Lina helped others and was nice to everyone. Her \textbf{teacher said} kindness is good.''} \newline
{\scriptsize\color{charcoal}\textit{$\rightarrow$ Authority-based validation.}} &
\footnotesize
\emph{``When Lina saw a new child sitting alone, she \textbf{invited them to join} her game. Soon they were laughing together, and Lina realized kindness can make anyone feel \textbf{welcome}.''} \newline
{\scriptsize\color{darkblue}\textbf{$\rightarrow$ Action-based inclusion.}} \\

\bottomrule
\end{tabular}
}
\vspace{-4mm} 
\end{table*}

\emph{\textbf{2. Supervised Neural Rankers:}} We train lightweight rankers~\cite{dehghani2017neuralranking,burges2005ranknet,liu2009learning} to approximate the Judge~B signal ($y_B$) over frozen embeddings: \textbf{Pointwise MLP:} Regresses \cite{pobrotyn2020context} directly to the scalar score $y_B$. \textbf{Pairwise RankNet:} Optimizes a logistic loss over preference pairs $(s_i \succ s_j)$ derived from $y_B$, capped at 800 pairs per query to balance epoch duration~\cite{burges2005ranknet}.

\emph{\textbf{3. Dense Retrieval Baseline:}} We evaluate the BGE-M3 (Distilled) model described in Section~\ref{sec:methodology}. As noted, this model is used off-the-shelf and is not fine-tuned on our dataset; it serves as a strong dense baseline.

\emph{\textbf{4. Oracle \& Upper Bounds:}} \textbf{Cross-Encoder (RoBERTa):} A full interaction-based model fine-tuned on $y_B$ supervision. This represents the high-compute upper bound of judge distillation~\cite{liu2019roberta}. \textbf{B-Score (Oracle):} Ranking candidates directly by the supervisor's score ($y_B$). This represents the theoretical limit of signal transfer under the two-judge framework.

\emph{\textbf{5. Late-Interaction Re-rankers:}} We evaluate late-interaction models that balance efficiency and expressivity. \textbf{ColBERTv2}~\cite{santhanam2021colbertv2} encodes queries and documents into token-level embeddings and aggregates relevance via MaxSim interactions. \textbf{RankZephyr}~\cite{pradeep2023rankzephyr} is a zero-shot listwise LLM re-ranker that directly scores full candidate lists without task-specific fine-tuning. Both models operate on the same fixed candidate pools and are evaluated under the same leakage-free Judge~A protocol.

\subsection{Evaluation Metrics}
We evaluate with standard query-level IR metrics using \emph{only} Judge~A  labels~\cite{jarvelin2002cumulated,chapelle2009expected,manning2008ir}. We report nDCG@5 and ERR@5. For nDCG, we use graded gain $g(r)=2^{r}-1$ with $r\in\{1,\dots,5\}$; IDCG is computed per query by sorting candidates by Judge~A grades. For ERR, we map grades to relevance probabilities as $R(r)=\frac{2^{r}-1}{2^{r_{\max}}}$ with $r_{\max}=5$~\cite{chapelle2009expected}.
We additionally report Kendall $\tau_b$ and Spearman $\rho$ computed per query between each method's scores and Judge~A grades, then macro-averaged across queries. Ties in grades are handled via average ranks~\cite{kendall1938rankcorrelation,spearman1904association}. For the Random baseline, correlations are expected to be near zero and are reported only as a sanity check rather than a stable comparison target.

\subsection{Results}
\label{sec:results}

We report leakage-free within-query ranking. All effectiveness numbers are computed exclusively using Judge~A labels. To ensure fair comparison, all methods rank the identical intersection pool $\mathcal{S}_q^{\cap}$ (defined in Section~\ref{sec:methodology}) per query.

Table~\ref{tab:main_results} reports leakage-free test performance at cutoff $k{=}5$ using 10 query-level splits (10 seeds), with 70 held-out test queries per seed. For each seed, we compute metrics per query and macro-average across the 70 test queries to obtain a seed-level score, and then report mean$\pm$std over 10 seeds. All results are computed on the fixed intersection pool $\mathcal{S}_q^{\cap}=(y_A \cap y_B)$ and query-level splits, so that all candidates for the same query $q$ are confined to a single split (train/dev/test) within each seed, preventing query-level leakage.

\emph{\textbf{Baselines vs.\ Judge-Driven Ranking:}}
BM25 improves over Random, but remains below judge-based signals, indicating that lexical matching captures some shallow cues yet is insufficient for culturally grounded normative selection (Table~\ref{tab:main_results})~\cite{robertson2009probabilistic}. Among classical IR baselines, performance is comparable to BM25 in nDCG@5, but they do not consistently improve effectiveness under our Judge~A metric, reinforcing that classical term/statistical matching is not tailored to normative cultural selection~\cite{amati2002dfr,zhai2004study}. Notably, distilling the cross-encoder supervision into the same BGE-M3 architecture yields a substantial further gain, outperforming the Judge-B cross-encoder and other learned rankers while preserving bi-encoder efficiency. In contrast, the distilled BGE-M3 bi-encoder yields a large improvement, indicating that teacher--student transfer can inject Judge~B supervision into an efficient dual-encoder scoring function. Directly sorting candidates by Judge~B rubric grades (B-Score) yields a large gain under leakage-free Judge~A evaluation and substantially higher within-query agreement with Judge~A, showing that rubric-based cultural judgments encode an ordering signal that transfers to an independent evaluator under strict separation.

\emph{\textbf{Learning to Rank from Judge~B Supervision:}} Neural rankers trained exclusively on Judge~B supervision transfer only modestly under leakage-free Judge~A evaluation. A listwise variant (ListMLE)~\cite{lan2014position} improves substantially over the pointwise and pairwise MLP rankers in effectiveness, but exhibits weak overall rank agreement with Judge~A. Despite this gain, ListMLE remains below the distilled bi-encoder and is also below directly sorting by Judge~B rubric grades (B-Score), indicating that listwise optimisation alone does not recover Judge~B transferable normative ordering without teacher-student transfer.

\emph{\textbf{Cross-Encoder (Teacher) Performance:}} A RoBERTa-large cross-encoder~\cite{liu2019roberta} trained on Judge~B supervision surprisingly substantially underperforms the distilled bi-encoder under leakage-free Judge~A evaluation. We attribute this counter-intuitive inversion to estimator overfitting: the high-capacity Cross-Encoder likely learns the specific idiosyncrasies and noise of the supervising judge ($J_B$), which do not transfer to the independent evaluator ($J_A$). In contrast, the distilled bi-encoder's bottlenecked architecture acts as a regulariser, forcing it to learn robust normative directions rather than judge-specific noise, thereby transferring the supervision more effectively than the teacher itself.

\emph{\textbf{Late-Interaction Re-ranking (ColBERTv2):}}
We additionally evaluate ColBERTv2 as a late-interaction re-ranker on the same per-query candidate pools (343 queries in total; 70 held-out test queries per seed).
ColBERTv2 achieves nDCG@5 $=0.635\pm0.058$ and ERR@5 $=0.707\pm0.043$, with within-query agreement of $\tau_b{=}0.209\pm0.035$ and $\rho{=}0.250\pm0.042$ under leakage-free Judge~A evaluation~\cite{santhanam2021colbertv2}.
We report ColBERT separately (Table~\ref{tab:colbert_results}) as it represents a distinct interaction rule from both bi-encoders and full cross-encoders: it consistently outperforms lexical and dense unsupervised baselines, and is competitive with interaction-based models under our Judge~A evaluation.

\emph{\textbf{Instruction-tuned LLM Re-ranking (RankZephyr):}}
We additionally evaluated RankZephyr (7B) \cite{pradeep2023rankzephyr}, a recent instruction-tuned ranking model, as a zero-shot re-ranker over the same per-query candidate pools. RankZephyr is evaluated with a fixed prompt template and deterministic decoding. Despite its strong performance on general web ranking benchmarks, RankZephyr (7B) achieves nDCG@5 $=0.507\pm0.047$ and ERR@5 $=0.621\pm0.049$ under Judge~A evaluation, underperforming Random and BM25 in our setting.
This suggests that generic instruction-following rankers do not reliably capture the fine-grained normative and culturally grounded distinctions required for within-query story selection, reinforcing the need for explicit judge-based supervision.

\emph{\textbf{Instruction-Tuned LLM Re-ranking (RankGPT-direct):}}
We evaluate RankGPT-direct (Mistral) as a zero-shot listwise re-ranker under the same leakage-free setting ($k{=}5$) and Judge~A-only evaluation \cite{sun2023instructiondistillation}. RankGPT achieves nDCG@5 $=0.559\pm0.025$ and ERR@5 $=0.662\pm0.018$, but underperforms Random/BM25 in nDCG and shows negative rank agreement with Judge~A. We therefore report RankGPT as an auxiliary diagnostic rather than a main baseline under our fixed-prompt, deterministic-decoding protocol.

\emph{\textbf{Rank-Correlation Sanity Check:}} We additionally report within-query rank agreement between each method’s scores and Judge~A labels using Kendall’s $\tau_b$ and Spearman’s $\rho$ (computed per query, then macro-averaged across test queries and seeds; undefined cases are excluded). We report $\tau_b$/$\rho$ only for methods that output real-valued scores comparable to JudgeA labels; lexical/random baselines are omitted in Table\ref{tab:main_results}. Correlation is most informative for methods that produce real-valued scores (dense retrievers, judge-based scoring, and learned rankers), as it reflects agreement beyond top-$k$ utility. B-Score exhibits substantially higher agreement with Judge~A ($\tau_b{=}0.297\pm0.035$, $\rho{=}0.326\pm0.038$), while the distilled BGE-M3 shows high agreement ($\tau_b{=}0.313\pm0.027$, $\rho{=}0.377\pm0.032$). 
For methods run with deterministic decoding, variation across seeds reflects different test-query sets rather than sampling noise. The lightweight neural rankers show near-zero average agreement (pointwise $\tau_b{=}0.016\pm0.017$, $\rho{=}0.020\pm0.021$; pairwise $\tau_b{=}0.089\pm0.010$, $\rho{=}0.116\pm0.012$), consistent with their limited gains in nDCG/ERR under leakage-free evaluation.

\emph{\textbf{Paired Significance:}} Table~\ref{tab:q70_significance_bge} shows that BM25, BGE-M3 (Distilled), and B-Score achieve statistically significant improvements over Random in nDCG@5 (and also in ERR@5 for BM25 and BGE-M3 (Distilled)). In contrast, Cross-Enc does not significantly improve over Random in nDCG@5 and exhibits a significant decrease in ERR@5. In head-to-head comparisons, B-Score significantly underperformed BGE-M3 (Distilled) (negative $\Delta$ in both nDCG@5 and ERR@5), and Cross-Enc significantly underperformed both BGE-M3 (Distilled) and B-Score in nDCG@5 and ERR@5.

Paired sign-flip permutation tests (50k permutations over 700 aligned query$\times$seed observations) indicate that BM25, BGE-M3 Distilled, and B-Score yield statistically reliable improvements over Random in nDCG@5. Additionally, BGE-M3 (Distilled) yields large and statistically significant gains over Random in nDCG@5 and ERR@5. Consistent with the main table, the lightweight pairwise model does not provide a reliable gain over strong dense retrieval, and significantly underperforms Random in ERR@5 despite no meaningful difference in nDCG@5.

\emph{\textbf{Sensitivity to Pool Size and Cutoff ($k$):}} We verify that these findings are not artefacts of candidate density or evaluation depth. Binning test queries (Table~\ref{tab:pool_sensitivity_main}) by intersection pool size (Small $\le 35$, Medium $36-50$, Large $>50$) reveals consistent ranking trends: BGE-M3 (Distilled) consistently outperforms Random across all bins (Table~\ref{tab:pool_sensitivity_main}). Furthermore, varying the cutoff $k \in \{1, 3, 5, 10\}$ shows no rank reversal among methods. The relative ordering among the non-neural baselines remains stable across cutoffs, with BGE-M3 (Distilled) consistently outperforming Random.

\emph{\textbf{Sensitivity to Supervision Scale: }} Table~\ref{tab:q70_ablation_frac} reports a training-fraction ablation for the \textit{pairwise} ranker trained on Judge~B supervision, following the main results framework and evaluated exclusively under Judge~A labels.
Performance is broadly stable across training fraction (TF)=0.25--0.75 (within-seed variance), with a clearer improvement at TF=1.00 under leakage-free Judge~A evaluation.

\emph{\textbf{External Validity (Moral Stories):}} To verify that our framework captures generalised normative reasoning beyond the generated NGR-33k corpus, we applied the same two-judge pipeline to the Moral Stories dataset ($N{=}500$)~\cite{emelin2021moral}. We treat the norm statement as the query and the action as the candidate. As shown in Table~\ref{tab:moral_external}, Judge~A and Judge~B achieve strong agreement ($\rho=0.653$) on this human-curated benchmark. This confirms that the signal transfer observed in our main results is driven by a shared, latent normative alignment between the evaluators rather than synthetic dataset artefacts. In contrast, an out-of-domain check on SSGEN ($N{=}500$) yielded negligible correlation ($\rho \approx 0.046$), effectively delineating the boundary conditions of the judges' normative alignment. The ``Hall of Mirrors'' \cite{hellrigel2025misalignment} risk is mitigated by the strong transfer performance on the human-curated Moral Stories dataset, which anchors the synthetic supervision signal to real-world normative distributions.

\subsection{Qualitative Analysis}
\label{sec:qualitative}

To demonstrate the impact of normative ranking, we present a side-by-side comparison of top-ranked candidates from the Dense Baseline (BGE-M3) and our Judge-Driven Supervisor (B-Score). As shown in Table~\ref{tab:qual_comparison}, the Judge-driven ranker systematically selects stories with concrete, culturally situated practices (highlighted in bold), whereas the dense baseline tends to retrieve globally plausible but culturally generic narratives.

\emph{\textbf{Discussion:}} The examples in Table~\ref{tab:qual_comparison} illustrate the Normative Feature Suppression \cite{yu2023improving} phenomenon discussed in Section~\ref{sec:discussion}. The dense retriever (BGE-M3) successfully matches the topic of the moral (e.g., waiting for mangoes $\approx$ patience), but misses the normative texture required by the query. In contrast, the supervision signal from Judge~B explicitly rewards enactment: patience is not just waiting, but waiting to prepare a Rangoli; kindness is not just ``being nice'', but breaking bread under a tree. This confirms that our leakage-free ranking framework successfully transfers these subtle, non-topical preferences to the final ranking model.

\begin{table}
\centering
\caption{Judge Agreement on the human-curated Moral Stories (Target) vs. SSGEN (Out-of-Distribution Control). (\textit{n.s.}: not significant, $p > 0.05$).}
\label{tab:moral_external}
\vspace{-0.05in}
\resizebox{0.95\columnwidth}{!}{
\setlength{\tabcolsep}{8pt}

\begin{tabular}{lc cc c}
\toprule
& & \multicolumn{2}{c}{\textbf{Judge Correlation}} & \\
\cmidrule(lr){3-4}
\textbf{Dataset} & \textbf{$N$} & \textbf{Spearman $\rho$} & \textbf{Kendall $\tau$} & \textbf{Sig.} \\
\midrule

\rowcolor{green!10} 
\textbf{Moral Stories} (Target) & 469 & $\mathbf{0.653}$ & $\mathbf{0.594}$ & $<10^{-45}$ \\

\rowcolor{red!5} 
SSGEN (Control) & 462 & $0.046$ & $0.045$ & n.s. \\
\bottomrule
\end{tabular}
}
\vspace{-5mm}
\end{table}

\section{Discussion}
\label{sec:discussion}

\emph{\textbf{Methodological Rigour---Evaluating without Circularity:}} Our framework addresses the systemic \emph{self-preference bias} and \emph{preference leakage} inherent in single-judge evaluations~\cite{koo2024benchmarkingcognitive,chen2024humansllmsjudgestudy,li2025preferenceleakage}. We rigorously enforce estimator separation ($J_A \neq J_B$) to prevent the model from overfitting to the idiosyncrasies of a specific evaluator. We acknowledge, however, that ``leakage-free'' refers to the experimental framework, not the pre-training distribution. Since Yi-1.5 and LLaMA-3 likely share training data (e.g., CommonCrawl), they inevitably share latent societal priors. Critically, our results on the OOD control dataset (SSGEN) show that this shared background is insufficient to produce agreement on its own ($\rho \approx 0.04$). High agreement only emerges when the underlying narrative contains valid normative structures (Moral Stories, $\rho \approx 0.65$). This confirms that while the judges share a ``worldview'', our framework successfully isolates the learning signal from the evaluation signal, preventing the circular inflation typical of self-rewarding loops.

\emph{\textbf{Direct vs. Distilled Learning:}} We observe a stark contrast between direct supervision and teacher-student distillation. Bi-encoders trained directly on Judge~B scores fail to improve over baselines (nDCG $\approx$ 0.575), suggesting a Normative Feature Suppression where cultural nuances are overshadowed by dominant topical components in the embedding space. However, the BGE-M3 (Distilled) model achieves nDCG 0.771, substantially outperforming the Cross-Encoder (0.577). This confirms that the bi-encoder architecture \emph{does} possess the capacity to encode cultural norms, but the optimisation landscape under raw judge labels is too noisy for it to navigate alone. The Cross-Encoder acts as a necessary denoising teacher. Although the Cross-Encoder itself overfits to the supervisor's idiosyncrasies (resulting in lower test performance), its soft logits provide a smoothed gradient signal. This allows the student bi-encoder to locate the suppressed normative manifold while its lower capacity naturally filters out the teacher's overfitting. This phenomenon is distinct from the semantic collapse described by \citet{nguyen2025milco}, as the features are recoverable via distillation.

\emph{\textbf{Cross-Encoder Distillation to Bi-Encoders:}}
Distilling supervision from a Cross-Encoder into BGE-M3 substantially improves bi-encoder performance, narrowing the gap while preserving efficiency. This indicates that strong teacher signals can partially inject normative structure into dense embeddings, though a residual gap remains due to the loss of token-level interactions.

\emph{\textbf{The Regularisation Hypothesis (Student > Teacher):}} A counter-intuitive finding of our study is the distilled bi-encoder (Student) significantly outperforming its cross-encoder supervisor (Teacher) under independent evaluation ($\Delta$ nDCG +0.194). In the context of noisy proxy supervision ($J_B$), we posit that the Cross-Encoder suffers from estimator overfitting: its high capacity allows it to model the non-robust, judge-specific artefacts (e.g., Judge B's lexical biases or prompt sensitivities) that do not transfer to the independent Judge A. Conversely, the bi-encoder's architecture imposes a dot-product bottleneck, forcing the model to project relevance into a lower-dimensional semantic space. This bottleneck effectively acts as a noise filter \cite{kaplun2022knowledge}, preventing the student from learning the high-frequency, non-generalizable quirks of the teacher. This phenomenon aligns with findings in noisy student training \cite{xie2019self} and weak supervision \cite{karamanolakis2021self}, where students outperform teachers \cite{li2024self} by smoothing the decision boundary and discarding label noise. Thus, distillation in this framework serves not merely as compression, but as a robust form of architectural regularisation, stripping away the evaluator variance to reveal the latent normative signal.

\section{Conclusions}
\label{sec:conclusion}
We presented a leakage-free ranking framework for culturally grounded GenIR that mitigates preference leakage via disjoint evaluators. By rigorously decoupling supervision from evaluation, we confirm that cultural grounding is a robust ranking signal that persists even when the teacher is removed. Experiments on the NGR-33k benchmark reveal that classical lexical baselines provide only marginal improvements over Random, while dense semantic retrieval (BGE-M3) yields substantially strong performance; however, even a distilled bi-encoder remains clearly behind judge-supervised and teacher--student-distilled ranking models under leakage-free evaluation. Moreover, distilling the Cross-Encoder into a fine-tuned BGE-M3 bi-encoder substantially narrows this gap, demonstrating that teacher-guided representation learning can recover a large portion of the normative signal while preserving retrieval efficiency. This demonstrates that high-capacity interaction modelling is required to discover the normative signal, but once discovered, it can be effectively compressed into efficient dense embeddings for deployment.

\bibliographystyle{ACM-Reference-Format}
\bibliography{references}
\end{document}